\documentclass[%
reprint,
superscriptaddress,
 amsmath,amssymb,
 prl,
]{revtex4-2}

\usepackage[american]{babel}
\usepackage{graphicx}
\usepackage{dcolumn}
\usepackage{bm}
\usepackage{color}
\usepackage{epstopdf}

\begin{document}


\title{Creep control in soft particle packings.}

\author{Joshua A. Dijksman}
 \email{joshua.dijksman@wur.nl}
 \affiliation{Physical Chemistry and Soft Matter, Wageningen University \& Research, Stippeneng 4, 6708 WE Wageningen, The Netherlands}
 
\author{Tom Mullin}
 \email{tom.mullin@maths.ox.ac.uk}
\affiliation{The Mathematical Institute and Linacre College,
 University of Oxford.OX2 6GG,U.K.
}


\date{\today}

\begin{abstract}
Granular packings display a wealth of mechanical features which are of widespread significance. One of these features is creep: the slow deformation under applied stress. Creep is common for many other amorphous materials such as many metals and polymers. The slow motion of creep is challenging to understand, probe and control. We probe the creep properties of packings of soft spheres with a sinking ball viscometer. We find that in our granular packings, creep persists up to large strains and has a power law form, with diffusive dynamics. The creep amplitude is exponentially dependent on both applied stress and the concentration of hydrogel, suggesting that a competition between driving and confinement determines the dynamics. Our results provide insights into the mechanical properties of soft solids and the scaling laws provide a clear benchmark for new theory that explains creep, and provide the tantalizing prospect that creep can be controlled by a boundary stress.
\end{abstract}

\maketitle


\maketitle

The mechanical properties of athermal particle packings are of fundamental and applied interest and have various non-trivial features. Many systems such as sand, foams, emulsions and other particulate media have a ``rigid'' phase that can bear a finite amount of stress~\cite{evans_concentration_1990, bartsch_effect_2002, hecke_jamming_2009, shao_role_2013, vlassopoulos_tunable_2014, coussot2014yield, basu2014rheology, villone_dynamics_2019, o2019jammed, shewan2021viscoelasticity}. However, the definition of ``rigid'' is not always clear, for example because of slow mechanical motion or \emph{creep} in thermally driven amorphous materials~\cite{andrade1910viscous}. Packings of inelastic particles might be considered rigid, yet they also display slow relaxation dynamics when \emph{deformation} is imposed, even in the absence of thermal fluctuations; they are considered to self-fluidize~\cite{PhysRevLett.81.2934,PhysRevLett.103.036001}. Alternatively, when \emph{stress} is imposed, granular packings also display very small magnitude logarithmic aging~\cite{darnige2011creep, deshpande2021perpetual}. Hence, the origin of creep in athermal packings is obscure.\\ 
In this letter we show that athermal soft sphere packings display  readily observed, large amplitude creep behavior when the applied stress  of the intruder and the packing characteristics are controlled systematically. Observed creep behavior is diffusive in time, robustly observed under different experimental conditions and depends exponentially on the applied intruder stress. Our work shows that creep behavior can be systematically studied, which opens the door to finding the new physics needed to understand the slow flow behavior of (a)thermal particle packings. Specifically, our work suggests that creep features are set by a balance of applied and confinement stress and originate in particle or contact level details.\\

\begin{figure}[!t]
\centering
    \includegraphics[width = 0.22\textwidth]{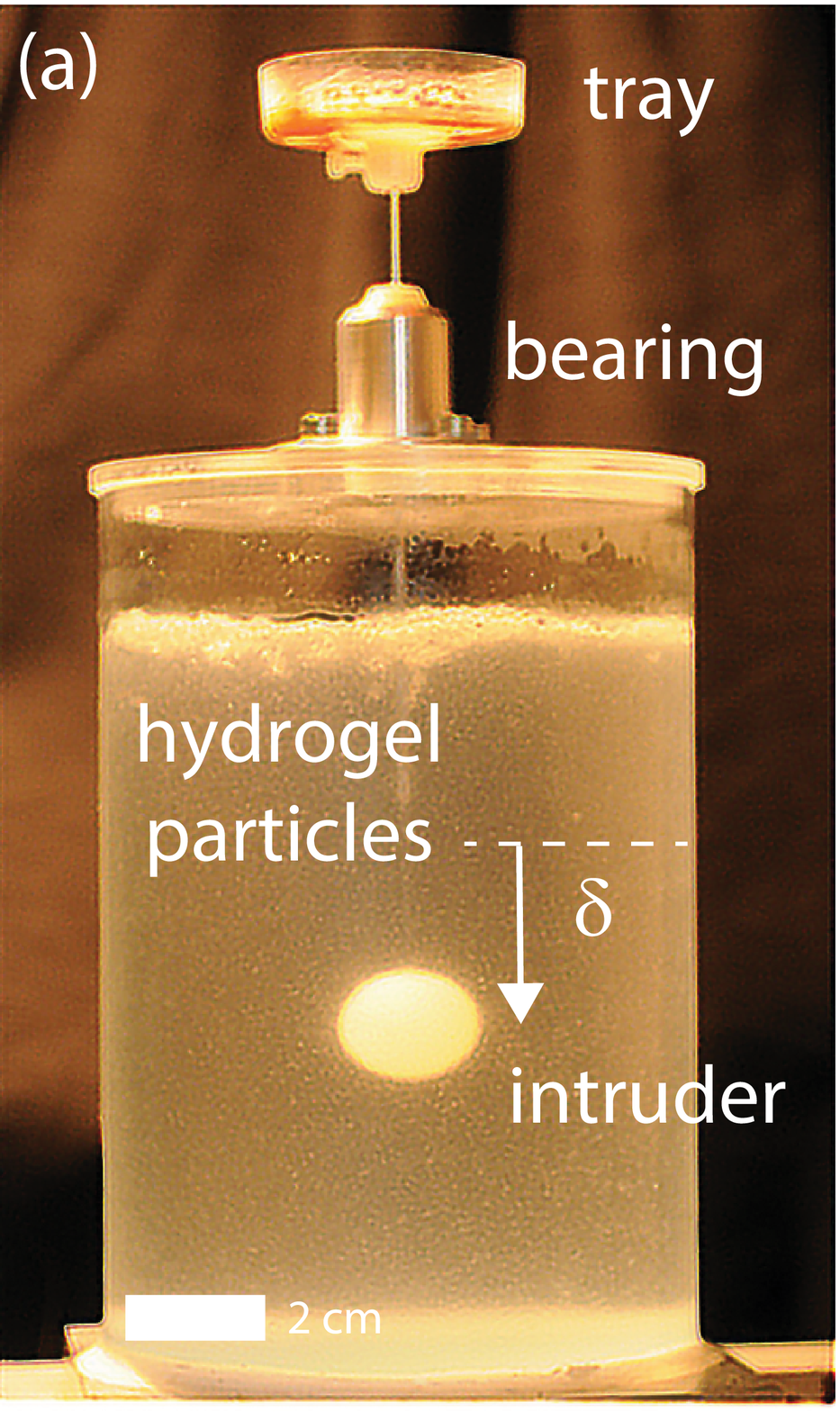}
    \includegraphics[viewport=5 5 280 530, clip=true, width = 0.20\textwidth]{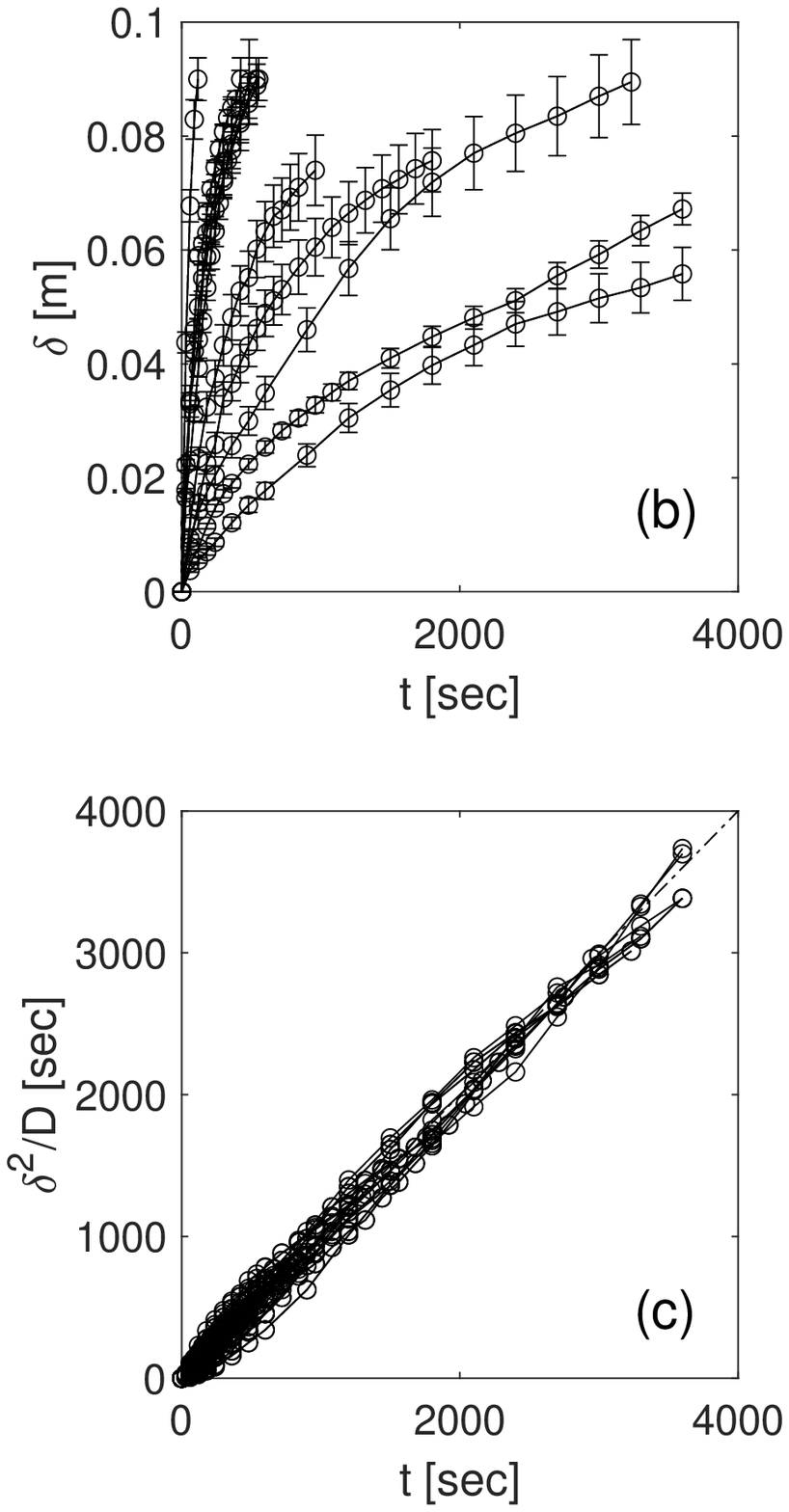}
\caption{(a) Photo of the experimental setup with various parts indicated. $\delta$ is the penetration depth measured from a reference point (dashed line) which is always below the surface of the packing. (b) Penetration depth $\delta$ as a function of time for various hydrogel concentrations and added weights. Error bars indicate measurement uncertainty on the position. (c) $\delta^2$ dynamics normalized with fitted $D$ values (see text) for all penetration tests. The dash-dotted line indicates a slope of unity.}
\label{fig:setup-examp}
\end{figure}

Microscopic details are always relevant to the understanding of the properties of soft solids. For example, it is known that constraining the various internal degrees of freedom of particles  leads to rigidity in idealized structures~\cite{calladine1978buckminster,hecke_jamming_2009}, e.g. where contacts proliferate. However, the properties of the rigid state also depend on the microscopic details of the particle contacts and adhesion, size, aspect ratio, boundary conditions, sample preparation and the way the rigidity of the assembly is probed. Controlling for all these variables is a huge challenge and a framework to understand rigidity is lacking in no small part due to the experimental challenges in systematically probing creep behavior. We therefore consider the classic viscosity test of a sinking  intruder of known shape and size and tunable effective density to probe the mechanical properties of packings of soft particles.. The medium we use is a dense packing of millimeter sized hydrogel spheres. In essence, we use the principles of a falling bob viscometer to study the properties of hydrogel particles packings. Such viscometers are commonly  used to obtain estimates of the viscosity of Newtonian fluids~\cite{faber1995fluid}, or non-Newtonian fluids~\cite{beris1985creeping,mckinley2002steady}. Viscoelastic fluids such as Boger fluids have been treated extensively, yet also ``simple'' yield stress fluids have been studied using sphere intrusion~\cite{gueslin2006flow,piau2007carbopol,PhysRevFluids.3.084303}. 

Unlike their microgel equivalents~\cite{plamper2017functional}, saturated millimeter sized soft hydrogel particles do not easily change volume under small pressures~\cite{cai2011mechanics,fengler2020desalination,louf2021under} and can be easily confined to a constant pressure or volume. They are known to have non-Newtonian flow behavior, even sans submersion~\cite{harth2020intermittent}. Hydrogels are virtually frictionless~\cite{workamp2019contact,cuccia2020pore} when fully immersed in water. Even so, a packing of frictionless spheres is known to have a yield stress~\cite{peyneau2008frictionless,workamp2019contact,PhysRevFluids.3.084303}, signalling its rigid feature. The macroscopic size of our system was selected to enable  accurate mechanical control. The intruder is much larger than the particle scale and thus provides a coarse-grained measure of the packing mechanics.  

\emph{Principle creep phenomenology ---} Spheres sinking into a non-Newtonian fluid typically  exhibit behaviour ranging from unsteady motion to arrest. Our experiments yield qualitatively different behaviors. (i) The sinking dynamics of the sinking sphere in the hydrogel particle packing does not stop~\cite{PhysRevLett.108.255701}; it creeps with reproducible power law time dependence. It is important to distinguish between the ``creeping flow'' limit of inertia free dynamics, and ``creep flow'', which is slow, inertia free motion that is nevertheless not steady due to ageing~\cite{struik1977physical}. (ii) The creep becomes exponentially dependent on the applied stress~\cite{hutchinson1995physical,lidon2017power} at a critical fraction of hydrogel material per unit volume $\rho_c$. (iii) Increasing the hydrogel concentration further, we find a transition from creep to ``creeping'' sinking behavior at a hydrogel concentration $\rho_l$ while retaining its exponential stress dependence. 

\textit{Experiments ---} The experimental apparatus is presented in Fig.~\ref{fig:setup-examp}a. We use an acrylic cylinder which contains a prepared packing of hydrogel spheres. The packing of hydrogel spheres can be adjusted by adding small measured amounts of dry hydrogel powder. An intruder is guided into the packing by an attached rod, that serves as a measure to track the penetration $\delta(t)$ into the medium. The rod also supports a tray, to which calibrated masses are added. At the start of each experiment the rod is held fixed using a clamp. The initial position of the intruder $\delta_0$ is with the top of the intruder positioned totally immersed, $5$~mm below the sample surface. The experiment is started by releasing the clamp. The total stress $\sigma$ exerted by the intruder on the packing is computed using the total  weight of intruder, rod, tray and added mass, divided by the cross section of the intruder. Note that this is a normal stress, computed with the cross section of the intruder. Details regarding the experiment can be found in the supplementary material [url], which describes the methods to prepare the samples and some other details and includes~\cite{oxfordwater21,ladenburg1907influence,derec2001rheology}. 

\begin{figure}[t]
\centering
\includegraphics[width=0.47\textwidth]{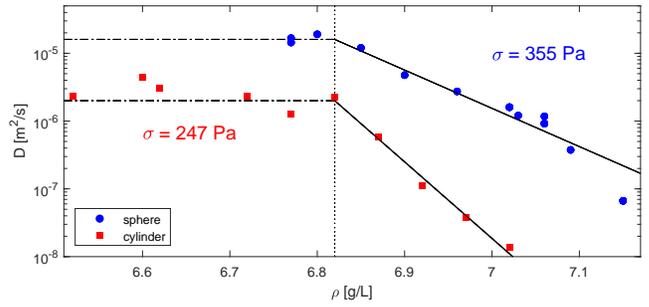}
\caption{At constant stress, we measure $D(\rho)$ for the 23~mm sphere (blue circles) and the 30~mm cylinder (red squares) at constant sinking stress $\sigma$ of having no weights in the tray. The dash-dotted line indicates regime of density-independent creep; the solid line indicates density dependent creep intrusion dynamics. The dotted line indicates $\rho_c$.}
\label{fig:D-rho}
\end{figure}

\textit{Creep features ---} The salient feature of the mechanics of the weak hydrogel particle packings is that they exhibit power law creep over a wide range of hydrogel concentrations and applied stress. Results for $\delta(t) = (D(\rho,\sigma)t)^{0.5}$ from sphere intrusion dynamics are shown in Fig.~\ref{fig:setup-examp}b for various $\sigma,\rho$. $D$ here is a diffusion-like constant with units of m$^2$/s that sets the penetration speed. The unsteady penetration dynamics is present in all experimental runs and persists for hours. We focus first on the sphere intrusion with no added weights in the tray which corresponds to a constant stress of approximately 360~Pa. The square root time dependence of $\delta$ is evidenced by plotting $\delta^2(t)$ normalised by the fitted $D(\rho,\sigma)$ as shown in Fig.~\ref{fig:setup-examp}c. In fact we obtain $D(\sigma, \rho)$ from a linear fit to $\delta^2(t)$.  All the data collapse onto a straight line with slope~1. See supplementary material for a logarithmic comparison of the same data. It should be noted that the distance over which creep persists is up to four times the intruder size.\\ 
The change in creep behavior at $\rho_c$ is readily observed in $D(\rho)$. Below a critical hydrogel concentration of 6.82~g/L, creep behavior exists with the same power law time dependence, yet $D$ is independent of $\rho$ and can be indicated by $D_0$. Above $\rho_c$, $D$ decays exponentially with $\rho$ as visible in Fig.~\ref{fig:D-rho}, where we can write:
\begin{equation}
    D = D_0\exp\left(-(\rho-\rho_c)/\rho_s\right).\label{eq:rho}
\end{equation} 
Here $\rho_s$ is a decay constant that we fitted to the data. We find that for the sphere, $\rho_s \approx 1/13$ g/L. This $D(\rho)$ behavior is also found for cylindrical intruders, for which $\delta(t)$ also has a square-root fit. The constant $D$ regime is more extensively surveyed with cylindrical intruders; for the 30~mm cylinder $\rho_s \approx 1/26$ g/L. Independent measurements with a larger cylindrical intruder show similar exponential $D(\rho)$ behavior (not shown).  Remarkably, we thus see that a packing of hydrogel spheres behaves as a soft solid that shows creep over the entire range of $\rho < \rho_l$ studied. The creep is reminiscent of thermally activated materials such as polymers and glasses but also found in harder particulate media~\cite{jaeger1989relaxation,d2003observing,dijksman2011jamming,darnige2011creep,hanotin2012vibration}.

\begin{figure}[!t]
\centering
\includegraphics[width=0.52\textwidth]{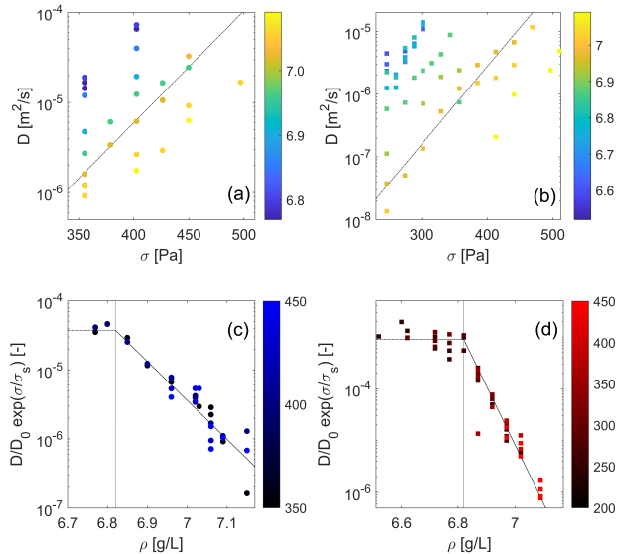}
\caption{Varying the intruder stress on the packing, (a,c) $D(\rho, \sigma)$ for the 23~mm sphere and (b,d) for the 30~mm cylinder. (a,b) show the exponential intruder stress dependence. The dash-dotted line is $D \propto \exp(\sigma/\sigma_s)$. Color scale indicates the hydrogel concentration. In (c,d) we rescale the data with Eq.~\ref{eq:stress}. Color scale indicates the applied stress $\sigma$ in Pa. Line styles are identical to Fig.~\ref{fig:D-rho}. }
\label{fig:D-sigma}
\end{figure}

\textit{Intruder stress dependence ---}  Increasing the applied stress $\sigma$ by adding weights to the tray increases the penetration speed. We find an exponential enhancement of $D_0$ with stress, as shown in Fig.~\ref{fig:D-sigma}a,b. In this exponential dependence, a stress scale $\sigma_s$ quantifies how creep changes with applied stress. For both spheres (panel a) and cylinders (panel b), $D$ increases exponentially with the same stress scale $\sigma_s \approx 35$Pa and indeed the data collapse of $D/D_0\exp(\sigma/\sigma_s)$ is excellent, especially above $\rho_c$. Normalizing $D$ by $D_0\exp(\sigma/\sigma_s)$ does not scale out the diffusion amplitude entirely. We therefore conclude that the overall prefactor $D_0$ can be described as
\begin{equation}
D_0 = D_{\sigma}\exp(\sigma/\sigma_s)\label{eq:stress}
\end{equation}
Here $D_{\sigma}$ is a geometry dependent factor with $D_{\sigma}\approx3.5\times10^{-3}$~m$^2$/s for the spheres and $D_{\sigma}\approx9\times10^{-4}$~m$^2$/s for the cylinders. As the sphere is only 1.5~times smaller than the sphere, we expect the remaining difference in $D_{\sigma}$ to arise from geometric effects: the flat face of the cylinder pushes small numbers of particles along, effectively changing its shape.

\textit{Interpretation ---}
The remarkable stress dependence observed by varying the effective mass and the hydrogel concentration suggests a mechanism connects the two effects. Indeed, the scaling for $D$ in the regime $\rho_c < \rho < \rho_l$ is
\begin{equation}
    D = D_{\sigma}\exp\left(\frac{\sigma}{\sigma_s}-\frac{(\rho-\rho_c)}{\rho_s}\right).\label{eq:D}
\end{equation} 
It seems more natural to see the diffusion constant as a balance of two competing stresses. First, $\rho_c$ is intruder shape independent, so it is a property of the packing. We observe that the fully hydrated, swollen hydrogel particles start to protrude through the water surface when the hydrogel concentration has reached $\rho_c$. We interpret $\rho_c$ as the hydrogel particle concentration at which the number of swollen hydrogel spheres fills the available volume of water. When dry hydrogel powder is added to an already completely filled liquid volume of water, particles swell to a hydrated state and protrusion increases. Surface tension then becomes relevant as the confining stress for the hydrogel particles. The surface tension stress is $2\gamma/d$ with $\gamma$ the surface tension and $d$ the average particle diameter, and yields about 100~Pa. This value is larger than the measured value for the geometry independent $\sigma_s$ but closer to this value than the other stress scale in the system: hydrostatic pressure. With 6g/L hydrogel, where the maximum hydrostatic pressure is $\Delta\rho gh \approx 12$~Pa with $h$ the packing height of about 20~cm. We conclude that creep speed appears to be set by the ratio of driving stress from intruder mass and the confining stress from the surface tension. Preliminary experiments suggest that adding a surfactant can increase $D$ at given $\rho$, but more work is needed to quantify these effects .\\ 

\emph{Origin of geometry dependent prefactor $D_{\sigma}$ ---} On dimensional grounds, in the creep regime, since $D_{\sigma}$ has the units of a diffusion constant m$^2$/s, this implies it will depend on a characteristic stress scale $\sigma$ of the intruder, on a density or concentration $\rho$ and a length scale $L$ via $D_{\sigma} \propto L\sqrt{\sigma/\rho}$. $\sigma_s$ is universal, hence a natural stress scale, the more so since the hydrostatic pressure scale is both too small and depth dependent and thus unlikely to produce the observed creep dynamics. While our experiment is limited in the exploration of different sizes, shapes and material densities, we  find that $\rho_s$ and $D_{\sigma}$ are intruder dependent and thus likely related. However, the value of $\rho_c, \rho_s$ requires careful interpretation, as the dry hydrogel weight per unit of volume of water is  a proxy for  the collective dynamics of the swollen hydrogel particles. Having established the values of $\rho$, we can estimate $L \approx D_{\sigma}\sqrt{\rho_s/\sigma_s}$ and find that $L \approx$ 1-2mm, which is equal to the particle size. The appearance of this length scale is consistent with the surface tension argument above, which would indeed include surface tension parameters in the effective prefactor $D_{\sigma}$. The relevance of surface tension for the creep behavior also provides an interpretation for the geometry dependent $\rho_s$: this constant is related to the amount of Reynolds dilatancy the packing of hydrogel particles experiences under the motion of the intruder. $\rho_s$ for the cylinder is thus also much smaller.\\ 
It is interesting to observe the creep and its exponential dependence on both $\rho$ and $\sigma$.  Preliminary  flow field visualizations rule out an emerging and compressing solid-like region underneath the intruder; the flow field during penetration is known to be limited around the sphere~\cite{gueslin2009sphere,PhysRevFluids.3.084303}. Moreover, creep is consistently observed for both intruder shapes. We postulate that the soft hydrogel packing acts as an athermal glass former~\cite{lyon2012polymer} in which \emph{any} motion of the intruder destabilizes the packing and induces fluidization, which gives rise to motion. The self-fluidization will be suppressed under larger confining pressure, or enhanced when the driving pressure is higher. We cannot probe the limit $\sigma \to 0$ so we are unable to verify whether the self-fluidization persists to the point where the material becomes a ``solid'', yet if Eq.~\ref{eq:stress} extends to infinitesimal stress levels, our observations indicate that frictionless, deformable particles allow minute mechanical fluctuations to liquefy a packing of jammed particles. The anti-thixotropy~\cite{larson2019review} or aging phenomenon observed fits remarkably well in the framework from Derec {\textit et al}~\cite{derec2001rheology}, with an aging exponent $\alpha = 3$ (see supplementary material).  

\begin{figure}[!t]
\centering

\includegraphics[width=0.47\textwidth]{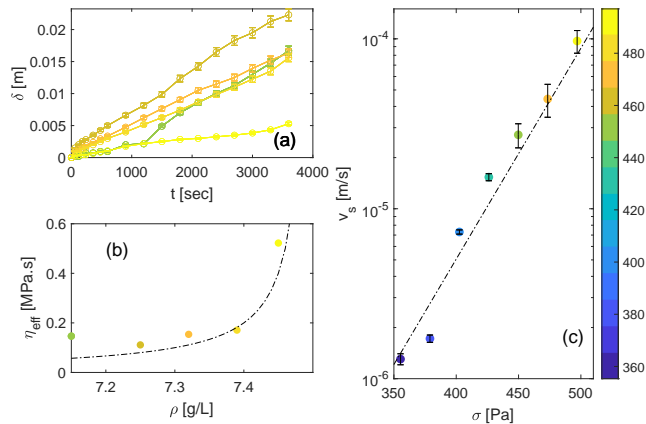}
\caption{(a) Above $\rho_l$, penetration depth $\delta$ as a function of time for various hydrogel concentrations at constant stress $\sigma = 358$~Pa. (b) the effective viscosity as derived from the sinking speed $v_s$. Color indicates $\rho$. The dash-dotted line shows a $\eta_{\rm eff} \propto 1/(7.5-\rho)$ divergence as guide to the eye. (c) $v_s$ as a function of stress $\sigma$. At fixed $\rho = 7.45$~g/L, $v_s$ is approximately exponential; the dash-dotted line is given by $v_s \propto \exp(\sigma/\sigma_s)$. $\sigma$ is indicated by the color scale, in Pa. Error bars indicate the 95\% confidence interval of the fit.}
\label{fig:linear}
\end{figure}

\textit{Robust $\sigma_s$ dependence in viscous sinking --- } The diffusive penetration dynamics does not extend to arbitrarily large hydrogel concentrations, but the stress dependent dynamics does. At a transition hydrogel concentration $\rho = 7.15$~g/L we  observe a mixed sinking behavior that cannot be described by unsteady creep with a single  exponent. Upon increasing $\rho > 7.15$~gr/L we enter a third regime. In this  regime, the sphere sinks at a constant speed. Typical $\delta(t)$ behavior is shown in Fig.~\ref{fig:linear}a and contrasts starkly with the creep behavior in Fig.~\ref{fig:setup-examp}b. Here we fit $\delta = v_st$ and using Stokes drag law we find that $v_s = 4R^2g\Delta\rho/18\eta_{\rm eff}$, with $R$ the intruder radius, $\eta_{\rm eff}$ the effective viscosity of the hydrogel packing and $\Delta\rho$ the density difference between the intruder + weights and the water-hydrogel mixture. The $\eta_{\rm eff}(\rho)$ dependence is weak but a slowdown with increasing hydrogel concentration can be observed, as can be seen in Fig.~\ref{fig:linear}b. In a simple fluid the speed of intrusion varies linearly with  added mass and hence $\sigma$, but the dependence on stress is significantly stronger, as is $D_{\sigma}$. We clarify this in Fig.~\ref{fig:linear}c where we show various $v_s(\sigma)$ at constant $\rho = 7.45$~g/L. Again, an exponential dependence of the effective viscosity on the stress can be seen. The reference line shown follows Eq.~\ref{eq:stress} and has the same $\sigma_s$ as used for describing $D$ below $\rho_l$ suggesting a robust stress dependence across regimes.

\textit{Conclusions ---} By varying the intruder stress and amount of hydrogel particles in a fixed volume of water, we find that creep in athermal soft particle packings can be systematically observed and depends exponentially on the intruder stress under a wide range of conditions. We find three regimes of sinking dynamics: (i) a regime with power law intrusion creep with an exponent of 0.5 for the penetration versus time dynamics, in which the hydrogel concentration does not affect the sinking speed prefactor. (ii) A second regime with the same power law intrusion creep in which the speed is exponentially dependent on the packing and the distance to a critical volume fraction. (iii) a regime of constant sinking rate. The observations are robust for different intruder shapes. We interpret the observed exponential stress dependence from a balance of driving stress and confinement pressure induced by surface tension and link the observed power law creep to a different type of aging than commonly found in logarithmic creep experiments. Our results provide a clear challenge for numerical and theoretical work. The observed exponential stress dependence provides a test for existing frameworks of particle packing mechanics and point towards a relevance of understanding particle contact level physics and boundary conditions on the fluctuations in granular packings to arrive at coarse grained level descriptions of disordered particulate media. Our extensive set of experimental results prompt several questions: which mechanism allows for the creep phenomenon? What sets the observed stress scale ? Is a boundary stress indeed responsible for the local flow behavior? Our data so provides a benchmark for perspectives proposed for mechanics in athermal packings~\cite{PhysRevLett.89.165501, PhysRevLett.104.165701, silbert2010jamming,fielding2020elastoviscoplastic} and will stimulate new experimental and theoretical work in the field.

\begin{acknowledgments}The experiments were initiated in the Observatory in the Mathematical Institute at Oxford University. The majority of the experiments were completed in TM's house during lockdown. TM's wife, Sylvia is thanked for her forebearance. TM is grateful to Dominic Vella and Lucie Domino for their support. The authors are also grateful to Keith Long who manufactured the apparatus. Chandan Shakya is thanked for carrying out preliminary experiments on the rheology and flow behavior of hydrogel suspensions. We thank Iker Zuriguel and Jorge Peixinho for providing useful feedback. JAD acknowledges funding from the European Union’s Horizon 2020 research and innovation program under the Marie Skłodowska Curie grant agreement No 812638.
\end{acknowledgments}

\bibliographystyle{unsrt}
\bibliography{hydrogel_sinking.bib}

\pagebreak

\emph{Supplemental Materials}\\
\textit{Experimental details ---} The main component of the experimental setup is a $200$~mm long extruded and machined acrylic cylinder which has a $130~\pm1$mm outer diameter and wall thickness  $3$~mm. A tight fitting lid prevents evaporation and contains a central PTFE bearing though which a rod passes that is connected to the intruders. The bottom $10$~mm of the rod is threaded to match the hole in the centre of the intruder. To increase the applied stress $\sigma$ on the packing, a $55$~mm diameter$\times10$mm high acrylic tray is attached to the rod, to which calibrated masses can be added. We use two different intruder shapes: a polypropylene sphere of $23$~mm diameter with density $913$~kg/m$^3$. Note that the polypropylene sphere connected to the rod and tray has an effective density higher than water. We also use plexiglas cylinders with a diameter of $30$~mm and $35$~mm and length $20$~mm and density $\rho_p=$1190~kg/m$^3$. For all sinking probes, the Ladenburg correction is small~\cite{ladenburg1907influence}. Aluminum rods attached to the intruders are $2$~mm diameter are long enough to allow for tens of millimeter of travel of the intruder. The weight of the rod was $1.74$~grams for the cylinders and $2.24$~grams for the sphere. The position of the intruder is measured using an traveling telescope with a vernier scale; timing is done using a digital stop watch. Readings are taken at one minute intervals. The error in position is  set by the speed $\dot{\delta}$ at the time of the reading.\\  

\textit{Packing preparation ---} The hydrogel packings are prepared by mixing a known weight of dry hydrogel powder (JRM Chemicals) with $2$~liters of Oxford tap water, which has a pH of about 8~\cite{oxfordwater21} and contains about 5~ppm sodium. Both pH and salinity affect hydrogel swelling characteristics~\cite{cai2011mechanics,fengler2020desalination}. The water is boiled three times to remove  dissolved gases; this method is an effective way of reducing the number of bubbles that might otherwise emerge in the sample. The water is cooled before mixing the dry hydrogel powder to create a low density  sample. All measurements are carried out at 19~$\pm~1~^o$C and the total liquid volume was also large enough to ensure temperature stability during an experiment. The mixture is stirred very slowly  and allowed to settle for periods ranging from $2$ to $4$~days beneath a plunger which covered the entire surface. The plunger has a $380$~gram weight attached to it so that each water/hydrogel sample was compressed to a dense packing by the same force before each run of the experiment, aiming for a protocol that ensures repeatability. It should be noted that the packing of particles always represents a settled bed. The frictionless nature of the particle interactions makes it very likely that the packing fraction is thus very close to random close packing at all times.\\ 
The hydrogel immersion in water all but removes the gravitational stress gradient, resulting in a graviational stresses of $<1000$~Pa in our $20$~cm tall sample as used in our experiment. We  measure the position of the inserted intruder using the cathetometer during each run of the experiment. We checked repeatability by redoing the experiment after stirring the sample and allowing it to settle. To change the sample after each run, we use a jeweller's balance to measure out small quantities (typically $\sim 0.1$~grams) of dry hydrogel particles which are then mixed into the sample. All  packings are thus be characterized by a total concentration $\rho$ grams of hydrogel per liter of water. Other intrusion experiments involve increasing the stress by  adding weights to the tray.\\   

\textit{Stress calculations ---} Note that when the intruder is totally submersed, we  use the effective density $\rho_{\rm eff}$ of the intruder to compute the stress; $\rho_{\rm eff}$ depends on the density of the intruder. The addition of hydrogels also affects the average density $\rho_{\rm eff}$, but this effect does not measurably affect the calculation of the stress. During the slow penetration of the intruder, the rod also slowly sinks into the hydrogel packing, gradually reducing the effective weight by $8~\mu$g/cm, which can be neglected on the total weight.\\

\begin{figure}[!h]
\centering
\includegraphics[width=0.4\textwidth]{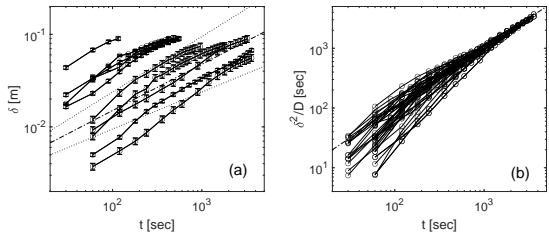}
\caption{Reproducing Fig.~1(b),1(c) from the main article on double-logarithmic scales. (a) Penetration depth $\delta$ as a function of time for various hydrogel concentrations and added weights. The dash-dotted line has a slope of 0.5. The dotted lines have slopes of 0.4 and 0.6. (b) $\delta^2$ dynamics normalized with fitted $D$ values for all penetration tests. The dash-dotted line indicates a slope of unity.}
\label{fig:setup-examp-dlog}
\end{figure}

\textit{Power law sinking rate ---} The power law nature of the creep dynamics can be established in more detail by examining $\delta(t)$ and its rescaled variant on double-logarithmic plots. In Fig.~\ref{fig:setup-examp-dlog} we show the same data as used for Fig.~1 of the main article on double-logarthimic scale. In Fig.~\ref{fig:setup-examp-dlog}a we include reference lines for different power law behavior to show that an exponent of 0.5 is very reasonable. The collapse shown in Fig.~\ref{fig:setup-examp-dlog}b indicates that during the first tens of seconds of the experiment, deviations may be observed from linearity, which are likely related to manual position measurement inaccuracies. The late time behavior, measured over several thousands of seconds, asymptotes to linearity on these rescaled axes.\\

\textit{Creep model ---} We use the framework for rheological constitutive equations for glassy and/or ageing materials described in~\cite{derec2001rheology,darnige2011creep}. Taking a standard scalar Maxwell model as a starting point, the stress dynamics is set by relaxation via the ``fluidity'' and shear modulus
\begin{equation}
    \dot{\sigma} = -f\sigma + G\dot{\gamma}\label{eq:sigmadot},
\end{equation}
where $G$ is the elastic modulus and fluidity $f$ is an inverse time associate with stress relaxation events. At the same time, $f(t)$ is described by the competition of an ageing process, increasing the characteristics relaxation time, and a rejuvenation process driven by shear. For the sinking ball experiments we can write that $\dot{\sigma} = 0$ and the observation is that $\delta(t) \propto t^{1/2}$ hence $\dot{\gamma} \propto t^{-1/2}$, although the scalar approach is not immediately justified. Substituting this, we find that $\sigma = t^{-1/2}G/f$ from which follows that $f \propto t^{-1/2}$ as the stress is held constant. With this in mind, the general form for $\dot{\gamma}$ as described by Derec \textit{et al} becomes of interest:
\begin{equation}
    \dot{f} = \left(r + u \left(\frac{\sigma f}{\dot{\gamma}}\right)^{\lambda}\frac{\dot{\gamma}^{\nu-\epsilon}}{f^{\nu}}\right)f^{\alpha}-vf^{\alpha+\beta},
\end{equation}
where all prefactors and exponents are positive constants whose magnitude determine the strength and type of aging and rejuvenation observed. For the hydrogel packing, we find $\dot{f} \propto t^{-3/2}$ hence one natural choice is $\alpha = 3$.
\end{document}